\documentclass[reprint, aps, pra, twocolumn, longbibliography, superscriptaddress, floatfix]{revtex4-2}

\usepackage{amsmath}
\usepackage{amssymb}
\usepackage{graphicx}
\usepackage{wasysym}
\usepackage{dsfont}
\usepackage{braket}
\usepackage{array}
\usepackage[ruled]{algorithm2e}
\usepackage[dvipsnames]{xcolor}
\usepackage{url}
\usepackage{units}
\usepackage{hyperref}
\usepackage{mathtools}
\hypersetup{
colorlinks = true,
urlcolor  = BrickRed,
linkcolor = RoyalBlue,
citecolor = ForestGreen
}

\SetAlgorithmName{Protocol}{protocol}{List of Protocols}

\begin{document}

\title{Device-Independent Quantum Key Distribution \\
Based on the Mermin-Peres Magic Square Game}

\author{Yi-Zheng Zhen}
\affiliation{Hefei National Research Center for Physical Sciences at the Microscale and School of Physical Sciences, University of Science and Technology of China, Hefei 230026, China}
\affiliation{Shanghai Research Center for Quantum Science and CAS Center for Excellence in Quantum Information and Quantum Physics, University of Science and Technology of China, Shanghai 201315, China}

\author{Yingqiu Mao}
\affiliation{Hefei National Research Center for Physical Sciences at the Microscale and School of Physical Sciences, University of Science and Technology of China, Hefei 230026, China}
\affiliation{Shanghai Research Center for Quantum Science and CAS Center for Excellence in Quantum Information and Quantum Physics, University of Science and Technology of China, Shanghai 201315, China}

\author{Yu-Zhe Zhang}
\affiliation{Hefei National Research Center for Physical Sciences at the Microscale and School of Physical Sciences, University of Science and Technology of China, Hefei 230026, China}
\affiliation{Shanghai Research Center for Quantum Science and CAS Center for Excellence in Quantum Information and Quantum Physics, University of Science and Technology of China, Shanghai 201315, China}

\author{Feihu Xu}
\email{feihuxu@ustc.edu.cn}
\affiliation{Hefei National Research Center for Physical Sciences at the Microscale and School of Physical Sciences, University of Science and Technology of China, Hefei 230026, China}
\affiliation{Shanghai Research Center for Quantum Science and CAS Center for Excellence in Quantum Information and Quantum Physics, University of Science and Technology of China, Shanghai 201315, China}
\affiliation{Hefei National Laboratory, University of Science and Technology of China, Hefei 230088, China}

\author{Barry C. Sanders}
\email{bsanders@ustc.edu.cn}
\affiliation{Hefei National Research Center for Physical Sciences at the Microscale and School of Physical Sciences, University of Science and Technology of China, Hefei 230026, China}
\affiliation{Shanghai Research Center for Quantum Science and CAS Center for Excellence in Quantum Information and Quantum Physics, University of Science and Technology of China, Shanghai 201315, China}
\affiliation{Institute for Quantum Science and Technology, University of Calgary, Alberta T2N 1N4, Canada}

\begin{abstract}
Device-independent quantum key distribution (DIQKD) is information-theoretically secure against adversaries who possess a scalable quantum computer and who have supplied malicious key-establishment systems;
however, the DIQKD key rate is currently too low.
Consequently, we devise a DIQKD scheme based on the quantum nonlocal Mermin-Peres magic square game:
our scheme asymptotically delivers DIQKD against collective attacks, even with noise.
%Compared with a family of DIQKD protocols based on the Clauser-Horne-Shimony-Holt game, \add{our scheme yields more secret keys given a number of executed nonlocal games,} when \add{both} the state visibility and detection efficiency are above certain thresholds.
%\add{
Our scheme outperforms DIQKD using the Clauser-Horne-Shimony-Holt game with respect to the number of game rounds,
albeit not number of entangled pairs,
provided that both state visibility and detection efficiency are high enough.
%}
\end{abstract}
\date{\today}

\maketitle

Device-independent quantum key distribution (DIQKD) enables distant parties to achieve quantum key distribution even with untrusted apparatuses~\cite{Ekert:91, MayersYao:98, BarrettEtAl:05}.
DIQKD provides information-theoretic security~\cite{VaziraniVidick:14, MillerShi:16, Arnon-FriedmanEtAl:18} against certain side-channel attacks that compromise the security of conventional quantum key distribution implementations~\cite{GisinEtAl:02, XuEtAl:20b, PortmannRenner:22}.
To achieve this security, DIQKD treats all devices that prepare, transmit, and measure information carriers as black boxes that could have been created by an adversary.
A nonlocality test~\cite{BrunnerEtAl:14} is typically executed by two communicating parties to estimate an adversary's possible knowledge about the generated data.
Based on the result of the test, the parties determine whether the data suffice to yield secure keys~\cite{VaziraniVidick:14, MillerShi:16, Arnon-FriedmanEtAl:18}.

However, as a sacrifice for high-level security, DIQKD yields a low key rate, as confirmed by recent experimental demonstrations~\cite{NadlingerEtAl:22, ZhangEtAl:22, LiuEtAl:22}.
For the potential use of DIQKD in practice, a high key-rate DIQKD protocol is demanding.
Here, we remedy this issue by employing the nonlocality test of a Mermin-Peres magic square game (MPG)~\cite{Mermin:90a, Peres:90} in DIQKD\@.
The MPG is a special nonlocal game whose quantum strategies allow two players to win with unit probability~\cite{BrassardEtAl:05, GisinEtAl:07}, thereby exceeding the winning probability of other nonlocal games such as the Clauser-Horne-Shimony-Holt (CHSH) game~\cite{ClauserEtAl:69}.
These remarkable features enable the MPG to yield a distinct DIQKD protocol from conventional protocols~\cite{PironioEtAl:09, AcinEtAl:07, WoodheadEtAl:21, SekatskiEtAl:21, Gonzales-UretaEtAl:21, PrimaatmajaEtAl:22}.

Here, we propose a DIQKD protocol based on the MPG
and prove security in the asymptotic case subject to collective attacks.
Adopting the technique proposed in Ref.~\cite{BrownEtAl:21a}, we numerically determine thresholds for state visibility and detection efficiency required by the protocol to generate secure keys.
We show that our MPG-based protocol generates a higher key rate, defined as the average number of secret bits generated in each instance of the protocol (namely, preparation, distribution, and measurement), compared to CHSH-based DIQKD protocols for certain parameter regimes.
%with the key rate defined as the average number of secret bits generated in each instance of the protocol, namely, preparation, distribution, and measurement.
%\add{
Precisely, we show that our MPG-based protocol demonstrates advantages if the state visibility exceeds 0.978 (with perfect detection) or if the detection efficiency exceeds 0.982
%\replace{0.952}{\textbf{0.982}}
(with a perfect source).
%}
Our results show the potential advantage of using more complex entangled states in implementing DIQKD\@.

\begin{figure}[hbtp]
 \centering
 \includegraphics[width=.95\columnwidth]{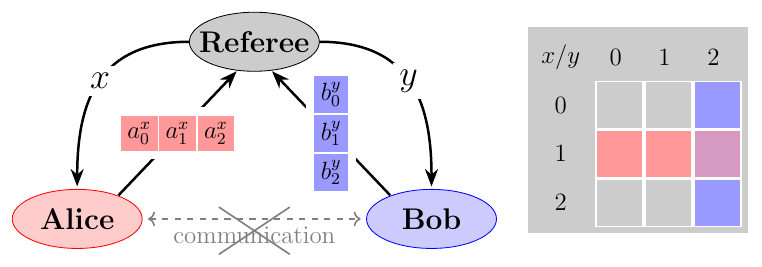}
 \caption{\label{fig:msg}%
 The Mermin-Peres magic square game.
 Two players, Alice and Bob, fill a $3\times3$ magic square over many rounds for both of them to win.
 In each round, a referee generates two random ``trits'' $x,y\in\{0,1,2\}$ and sends row index $x$ to Alice and column index $y$ to Bob.
 Alice and Bob then reply to the referee with a row $[a^x_0,a^x_1,a^x_2]$ and a column $[b^y_0,b^y_1,b^y_2]^\textsf{T}$, respectively, where 
 all $a^x_{i},b^y_{j}$ for $i,j\in\{0,1,2\}$ are bits that satisfy $\oplus_{i}a^x_{i}=0$ and $\oplus_{j}b^y_{j}=1$.
 The winning condition is that Alice and Bob share the same value for the overlapped grid, i.e., $a^x_{i=y}=b^y_{j=x}$.
 During the game, Alice and Bob are forbidden to communicate with each other.}
\end{figure}

{\em MPG-based DIQKD protocol.}---%%
Alice and Bob play the MPG~\cite{Mermin:90a, Peres:90, BrassardEtAl:05},
which is depicted and explained in Fig.~\ref{fig:msg}.
After the game, the referee decides whether Alice and Bob win or not according to the average winning probability
\begin{equation}
\omega=\sum_{x,y}\pi\left(x,y\right)P\left(a_{y}^x=b_{x}^y|x,y\right).\label{eq:msg-payoff-val}
\end{equation}
Here, $\pi(x,y)$ is the probability of distributing index pair $(x,y)$,
and $P(a_{y}^x=b_{x}^y|x,y)$ is the winning probability of Alice and Bob with respect to $(x,y)$.

Throughout, we employ the unbiased MPG: $\pi(x,y)=\nicefrac19$.
When using classical strategies (see one example in \S IA of Supplemental Material~\footnote{%
See Supplemental Material for the classical and quantum strategies for the MPG, the security of MPG-based protocol beyond the ideal case, and the CHSH-based DIQKD protocols.}%
), Alice and Bob's average winning probability is at most $\nicefrac89$~\cite{Cabello:01, GisinEtAl:07}.
As classical strategies are equivalent to local hidden variables, $\omega\leqslant\nicefrac89$ is actually a Bell inequality;
some quantum strategies violate this inequality~\cite{Cabello:01, GisinEtAl:07}.
Here, we denote a quantum strategy as $(\rho,{\cal M})$ and its average winning probability as $\omega(\rho,{\cal M})$, where $\rho$ is the distributed quantum state and ${\cal M}$ is the set of Alice's and Bob's quantum measurements used to generate the outputs.
In particular, when the state is two pairs of maximally entangled qubits
\begin{equation}
    \Psi_2
    =\Psi_{\text{A}_1\text{B}_1}^+
    \otimes\Psi_{\text{A}_2\text{B}_2}^+,\quad
    \Psi^+\coloneqq\frac{\Ket{00+11}\Bra{00+11}}{2},
\label{eq:two-ebits}
\end{equation}
a measurement set ${\cal M}_\text{opt}$ (details in \S IB of Supplemental Material~\cite{Note1}) 
exists such that Alice and Bob will win the MPG:
$\omega(\Psi_2,{\cal M}_\text{opt})=1$
with optimal quantum strategy $(\Psi_2, {\cal M}_\text{opt})$.

Crucially, the MPG certifies whether the outputs are correlated in every input pair.
This feature allows us to design a DIQKD protocol, as introduced in Protocol~\ref{alg:msg-diqkd-protocol}.
In this protocol, two communication parties,
also termed Alice and Bob, 
initially generate data by playing the MPG\@.
They announce their inputs and record the overlapped bits.
To estimate parameters, Alice communicates with Bob which part of the bits serves as raw keys with the remaining part of the bits announced to play the MPG\@.
If the average winning probability estimated from the announced data is less than an expected value $\omega_\text{exp}$, they abort the protocol;
otherwise, they perform data reconciliation on raw keys to obtain the final keys.
\begin{algorithm}[hbt]
\caption{\label{alg:msg-diqkd-protocol}The MPG-based DIQKD protocol}
\begin{description}
 \item [Input] $N$---number of rounds,\\ $\omega_\text{exp}$---expected winning probability of the MPG\@.
 \item [Output] $\boldsymbol{K}_{\text{A}}$---Alice's final key, $\boldsymbol{K}_\text{B}$---Bob's final key.
\end{description}\vspace{1mm}
\begin{description}
 \item [Data generation]
 In each round $n\in[N]=\{1,\dots,N\}$, Alice and Bob independently pick $x_n,y_n\in\{0,1,2\}$, uniformly at random.
 They inject $x_n$ and $y_n$ to their devices and record the outputs $[a^{x_n}_0,a^{x_n}_1,a^{x_n}_2=a^{x_n}_0\oplus a^{x_n}_1]$ for Alice and $[b^{y_n}_0,b^{y_n}_1, b^{y_n}_2=b^{y_n}_0\oplus b^{y_n}_1\oplus1]$ for Bob, respectively.
 \item [Announcement] Alice and Bob announce their inputs $\{x_n\}$ and $\{y_n\}$.
 They keep the bits $\{a^{x_n}_{y_n}\}$ and $\{b^{y_n}_{x_n}\}$, respectively.
 \item [Parameter estimation] Alice picks a random index subset $[K]\subsetneq[N]$ with a length $\gamma N$ and communicates $[K]$ with Bob.
 They use the bits with indexes in $[K]$ as raw keys $\boldsymbol{A}$ and $\boldsymbol{B}$, respectively, and announce the remaining bits, based on which they estimate the average winning probability $\omega$ of the MPG\@.
 If $\omega < \omega_\text{exp}$, they abort the protocol;
 otherwise, they proceed.
 \item [Data reconciliation] Alice and Bob apply error correction and privacy amplification on the raw keys $\boldsymbol{A}$ and $\boldsymbol{B}$ to obtain final secure keys $\boldsymbol{K}_\text{A}$ and $\boldsymbol{K}_\text{B}$, respectively.
\end{description}
\end{algorithm}

{\em Security analysis.}---%%
To prove security for a DIQKD protocol, one needs to consider general adversary attacks and finite-data effect~\cite{VaziraniVidick:14, MillerShi:16, Arnon-FriedmanEtAl:18}.
On the other hand, one can also temporarily consider weak security where adversary attacks are independent identically distributed (i.i.d.) collective and where the number of rounds $N$ is infinite (asymptotic scenario), followed by extending the weak security to the general scenario~\cite{Arnon-FriedmanEtAl:18, Arnon-FriedmanEtAl:19, TanEtAl:21c}.
Here, we analyze the weak security of Protocol~\ref{alg:msg-diqkd-protocol}.
%the full security of the protocol can be proven following procedures in~\cite{Arnon-FriedmanEtAl:19, TanEtAl:21c}.

Suppose that an MPG-based DIQKD protocol with a pre-determined $\omega_\text{exp}$ is successfully implemented.
We analyze if and how much secure key can be established for the case of i.i.d.\ collective attacks in the asymptotic scenario.
Because DIQKD assumes the correctness and completeness of quantum theory,  data generated in the protocol can be described by quantum measurements on a quantum state.
As all quantum devices are untrusted, the precise quantum state and quantum measurements are unknown.
Nevertheless, the assumption of i.i.d.\ collective attacks allows one to suppose that~\cite{PironioEtAl:09},
in each round, the adversary Eve produces a quantum state
$\psi_\text{ABE}$~\cite{PironioEtAl:09, Arnon-FriedmanEtAl:18} and distributes it to Alice and Bob.

Measurements in the protocol,
without losing generality, can always be described by the sets of projector-valued measures $\set{M_{a_0a_1}^{x}}_x$ for Alice and $\set{N_{b_0b_1}^{y}}_y$ for Bob, respectively.
Here,~$x$ and $y$ denote the inputs (i.e., measurement settings)
while $a_0a_1$ and $b_0b_1$ are bits representing Alice's and Bob's outputs,
\begin{equation}
[a^x_0,a^x_1, a^x_2=a^x_0\oplus a^x_1],\,[b^y_0, b^y_1, b^y_2=b^y_0\oplus b^y_1 \oplus 1]^\textsf{T},
\end{equation}
respectively.
For
\begin{equation}
\rho\coloneqq\text{tr}_\text{E}[\psi_\text{ABE}],\,
{\cal M}\coloneqq\Set{ \{M_{a_0a_1}^{x}\}_x,\{N_{b_0b_1}^{y}\}_y},
\end{equation}
$(\rho,{\cal M})$ is evidently a quantum strategy for the MPG.
The only constraint on $(\rho,{\cal M})$ is that the protocol is not aborted;
i.e., $\omega(\rho,{\cal M})\geqslant \omega_\text{exp}$.

An essential feature of the protocol is that raw keys are generated by all $(x,y)$ pairs of quantum measurements in ${\cal M}$ on the state $\psi_\text{ABE}$.
For each input pair $(x,y)$, whatever the precise forms of $\{M_{a_{0}b_{1}}^x\}$ and $\{N_{b_{0}b_{1}}^y\}$ are, the quantum state after the measurement can always be expressed as a classical-classical-quantum state.
\begin{equation}\label{eq:ccqstate}
\tau_{xy}:=\sum_{a^x_{y},b^y_{x}}\Ket{a^x_y}\Bra{a^x_y}_\text{A}\otimes\Ket{b^y_x}\Bra{b^y_x}_\text{B}\otimes \hat{\phi}_\text{E}\left(a^x_{y}b^y_{x}\right),
\end{equation}
where $a^x_{y}, b^y_{x}$ are the raw keys (bits from the overlapping grid of Alice's row and Bob's column) and $\hat{\phi}_\text{E}\left(a^x_{y}b^y_{x}\right)$ is the unnormalized quantum state characterizing Eve's knowledge of the raw keys.

An appropriate one-way data-reconciliation protocol always exists such that a certain amount of secure keys can be processed from the raw keys while eliminating Eve's information~\cite{DevetakWinter:05}.
As a result, for each input pair $(x,y)$, at least a ratio $\min\{0, r(\tau_{xy}) \}$ of the raw keys will remain as final keys, where
%In our protocol, this leads to that at least a ratio $\min\{0, r(\tau_{xy}) \}$ of the raw keys that will remain as final keys, for each input pair $(x,y)$, where
\begin{equation}\label{eq:Devetak-Winter}
 r\left(\tau_{xy}\right) \coloneqq H(\boldsymbol{A}|E)_{\tau_{xy}}-H(\boldsymbol{A}|\boldsymbol{B})_{\tau_{xy}}.
\end{equation}
Here, $H(~|~)$ denotes the conditional von Neumann entropy, and the bold symbols $\boldsymbol{A}$ and $\boldsymbol{B}$ imply that Alice and Bob's states are in classical states, representing the raw keys $\boldsymbol{A}$ and $\boldsymbol{B}$, respectively.

Finally, the security of the protocol should take all possible $\psi_\text{ABE}$ into account, which implies that the final keys correspond to the worst case of all quantum strategies $(\rho,{\cal M})$.
We define the key rate of DIQKD as the ratio of the length of final keys and the number of {\em all} rounds.
In the asymptotic limit $N\to\infty$, the key rate can be expressed as
\begin{equation}
R=\min\Set{0,\frac{\gamma}{9}\inf_{\left(\rho,{\cal M}\right)}\sum_{xy}r\left(\tau_{xy}\right)} ,
\label{eq:key-rate-msg}
\end{equation}
subject to
\begin{equation}
\omega\left(\rho,{\cal M}\right)\geqslant\omega_\text{exp}.
\label{eq:key-rate-msg-constr}
\end{equation}
Here, the coefficient $\nicefrac\gamma9$ comes from the fact that each input pair $(x,y)$ occurs with a probability $\nicefrac19$ while a ratio $\gamma$ of the rounds is used as raw keys.
The key rate defined here characterizes how many secure keys can be generated given the number of experimental rounds, which is different from the definition used in some literature where the test rounds are excluded~\cite{PrimaatmajaEtAl:22}.

To further prove that the protocol can produce correct and secure keys, we need to show that the key rate~$R$ has positive values when implementing the protocol using certain quantum strategies.
We first consider an ideal case, namely, implementing the protocol with the optimal quantum strategy $(\Psi_2,{\cal M}_\text{opt})$ and setting $\omega_\text{exp}=1$.
As the inputs are unbiased and randomly picked, the test rounds and the key rounds have the same correlations, both of which yield $\omega=1$, so the protocol will not be aborted.
Meanwhile, Alice and Bob's raw keys are uniformly distributed and perfectly correlated (see details in \S IB of Supplemental Material~\cite{Note1}).
We have $H(\boldsymbol{A}|\boldsymbol{B})_{\tau_{xy}}=0$ for any $(x,y)$.
Furthermore, $\omega=1$ in the MPG can self-test two singlets~\cite{WuEtAl:16};
i.e., the unknown quantum state must be locally isometric to two pairs of maximally entangled 2-qubit states.
Such states cannot be correlated with a third party because of entanglement monogamy~\cite{Terhal:04}.
Combining with the fact that $\boldsymbol{A}$ is uniformly random for all $\tau_{xy}$, we have $H(\boldsymbol{A}|E)_{\tau_{xy}}=1$, indicating that the adversary has no information on $\boldsymbol{A}$.
As a result, we obtain $R=\gamma$ for this ideal case, which shows that the protocol can indeed produce secure keys.

For non-ideal cases when general quantum strategies are adopted or when noises are involved, we can bound $R$ via the recently developed technique of quasi-relative entropy~\cite{BrownEtAl:21a}.
Precisely, a lower bound for $H(\boldsymbol{A}|E)_{\tau_{xy}}$ in Eq.~\eqref{eq:Devetak-Winter} can be derived as~\cite{BrownEtAl:21a}
\begin{equation}\label{eq:HAgElowerbound}
H\left(\boldsymbol{A}|E\right)_{\tau_{xy}}\geqslant c_{m}+\sum_{k=1}^{m-1}c_{k}\min\braket{\psi|G_{k}\left(x,y\right)|\psi},    
\end{equation}
where $c_{m}=\sum_{k=1}^{m-1}c_{k}$ and $c_{k}=w_{k}/(t_{k}\ln2)$, with $\{(t_{k},w_{k})|k=1,\dots,m\}$ a set of~$m$ nodes and weights of the Gauss-Radau quadrature, and $G_k(x,y)$ is defined as
\begin{align}\label{eq:Gk}
    G_{k}\left(x,y\right) &\coloneqq
    \sum_{a_{y}\in\left\{ 0,1\right\} }\left\{ \Pi_{a_{y}}^{x}\left[Z_{a_{y}}+Z_{a_{y}}^{\dagger} + \left(1-t_{k}\right)Z_{a_{y}}^{\dagger}Z_{a_{y}}\right]\right.\nonumber\\ 
    & \hspace{2cm} \left.+t_{k}Z_{a_{y}}Z_{a_{y}}^{\dagger}\right\},
\end{align}
with $Z_{a_y}$ an arbitrary operator and $\Pi_{a_{y}}^x$ the projector-valued measure corresponding to Alice's input~$x$ and output~$a^x_{y}$.
Combining with Eqs.~\eqref{eq:key-rate-msg} and~\eqref{eq:key-rate-msg-constr}, the minimization in Eq.~\eqref{eq:HAgElowerbound} is taken over all possible pure states $\ket{\psi}=\ket{\psi}_{\text{ABE}}$ and measurement strategies ${\cal M} = \set{\{M_{a_{0}a_{1}}^{x}\}_{x}, \{N_{b_{0}b_{1}}^{y}\}_{y}}$ subject to
\begin{subequations}\label{eq:SDP}
\begin{align}
    \omega\left(\rho,{\cal M}\right) &\geqslant  \omega_\text{exp},
    \quad \rho = \text{tr}_{\text{E}}\left[\ket{\psi}\bra{\psi}_{\text{ABE}}\right],\\ 
    \Pi_{a_{y}=0}^{x} &= 
        \begin{cases}
            M_{00}^{x}+M_{01}^{x} & \text{if }y=0\\
            M_{00}^{x}+M_{10}^{x} & \text{if }y=1\\
            M_{00}^{x}+M_{11}^{x} & \text{if }y=2
        \end{cases},\\
    \Pi_{a_{y}=1}^{x} &= \mathds1-\Pi_{a_{y}=0}^{x},\\ 
    0 &= \left[M_{a_{0}a_{1}}^{x^{\prime}},N_{b_{0}b_{1}}^{y^{\prime}}\right],\\
    0 &= \left[M_{a_{0}a_{1}}^{x^{\prime}},Z_{a_{y}}\right] = \left[Z_{a_{y}},N_{b_{0}b_{1}}^{y^{\prime}}\right],\\
    &\quad \forall a_{0,1,y}, b_{0,1}\in\{0,1\}, \forall x^{\prime},y^{\prime}\in\{0,1,2\}.
\end{align}
\end{subequations}
This constrained minimization can be resolved via the Navascués-Pironio-Acín (NPA) hierarchy~\cite{NavascuesEtAl:07}, which is numerically computable via solving a semidefinite program~\footnote{The $\mathsf{PYTHON}$ code for obtaining the key-rate bound can be found in~\url{https://github.com/YizhengZhen/Code_DIQKD_MPG}}.
In \S II of Supplemental Material~\cite{Note1}, we numerically show that $H(\boldsymbol{A}|E)_{\tau_{xy}}$ has a positive lower bound if $\omega_\text{exp}>0.9575$%
%\replace{0.915}{\textbf{0.9575}}
, which implies that the protocol can produce secure keys if, in the end, Eq.~\eqref{eq:key-rate-msg} has a positive value.

{\em Noise tolerance.}---%%
Consider the case where the optimal quantum strategy $(\Psi_2, {\cal M}_\text{opt})$ is supposed to be used to implement the MPG-based protocol.
Here, we characterize the performance of the protocol under two types of noise.
For the first type, we consider the imprecise preparation of $\Psi_2$ such that each qubit may be mixed with some white noise. 
The distributed state becomes $\rho_{\nu}\otimes\rho_{\nu}$ before the detection, where $\rho_{\nu}=\nu\Psi^++(1-\nu)\mathds1/2$ and $\nu$ is the state visibility.
For the second type, we consider the noise led by non-click events in measurements.
Such non-click events are caused by the loss of the state in transmission or the inefficiency of the detector, and cannot be sifted out from the data (otherwise, it may open detection loopholes~\cite{BrunnerEtAl:14, NadlingerEtAl:22, ZhangEtAl:22, LiuEtAl:22}).
Instead, Alice and Bob must assign an output to the non-click events.
We consider the following procedure: in each round, unless Alice (or Bob) successfully measured her (or his) state, Alice (or Bob) will output values according to a deterministic strategy of the MPG (Table I in \S IA of Supplemental Material~\cite{Note1}).
The detection efficiency on each side is assumed identical and denoted as $\eta$.

\begin{figure}[hbtp]
\centering
\includegraphics[width=0.95\columnwidth]{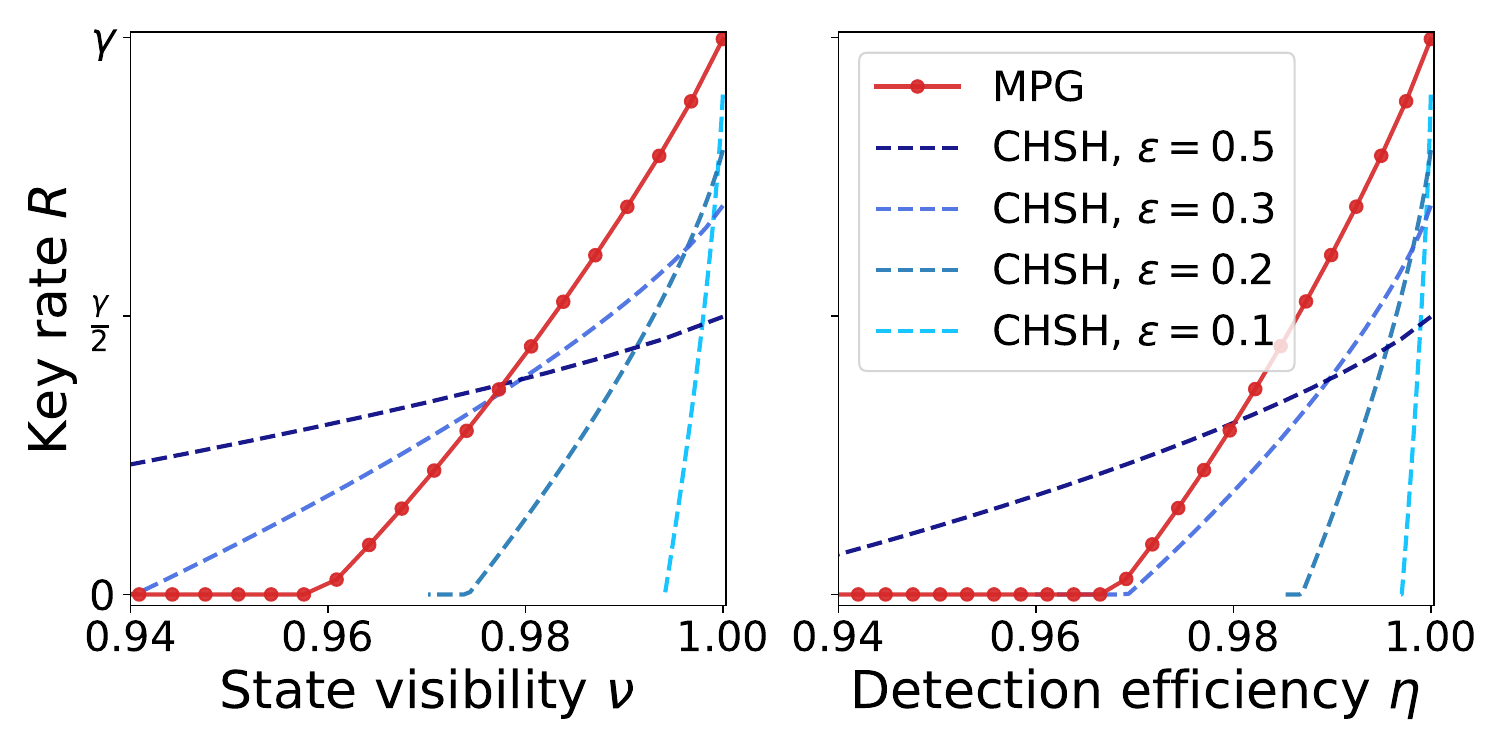}
\caption{\label{fig:key-rate-visibility-efficiency}%
    Noise tolerance of the MPG-based protocol (red solid line) against the state visibility $\nu$ (left) and the detection efficiency $\eta$ (right).
    %\replace{(right)}{\textbf{(right)}}.
    Results of the CHSH-based protocols are plotted as blue dashed lines for comparison.  $\varepsilon$ represents the probability of Alice picking $x=1$ in CHSH-based protocols.}
\end{figure}

We consider two cases where only the white noise is involved and where only the detection inefficiency is involved.
For each value of $\nu$ or $\eta$, we select $\omega_\text{exp}$ such that the produced data can pass the parameter estimation in the protocol.
The results are shown in Fig.~\ref{fig:key-rate-visibility-efficiency}.
In the figure, the red solid line is the key-rate lower bound of the MPG-based protocol, which is obtained by solving the semidefinite programming problem combining Eqs.~\eqref{eq:Devetak-Winter}-\eqref{eq:SDP}.
For the calculation, we set the NPA hierarchy as 2 and the number of nodes in the Gauss-Radau quadrature as 16.
From the figure, we observe that the key rate decreases when the state visibility or detection efficiency becomes smaller.
It shows that the MPG-based protocol can produce a positive key rate if the state visibility $\nu > 0.959$ or if the detection efficiency $\eta > 0.969$.
%\replace{0.942}{\textbf{0.969}}.

{\em Overcoming the key rate of CHSH-based protocols.}---%%
A prominent feature of the MPG-based protocol is that outputs of every input pair can be used to generate secure keys.
As a result, all the outputs except that used in the estimation are collected as raw keys.
This feature may enable the MPG-based protocol to yield a higher key rate than conventional DIQKD protocols.
Particularly, if the optimal quantum strategy of the MPG can be faithfully implemented, the key rate $R_\text{opt}=\gamma$.
It is actually the maximal key rate that any DIQKD protocol can achieve.
As we will show, the MPG-based protocol outperforms a variety of CHSH-based protocols for certain regions of noise parameters.

In a standard CHSH-based protocol~\cite{AcinEtAl:07,PironioEtAl:09}, Alice and Bob usually have inputs $x\in\{0,1\}$ and $y\in\{0,1,2\}$, respectively.
To generate the data, Alice picks~$x$ uniformly at random while Bob picks $y\in\{0,1,2\}$ according to probabilities $(1-\gamma)/2, (1-\gamma)/2, \gamma$, respectively, and they input~$x$ and $y$ into their local device and record the outputs $a,b\in\{0,1\}$.
After obtaining all the data, they announce the inputs and select the outputs corresponding to the input pair $(x,y)\in\{0,1\}^2$ to play the CHSH game.
If the average winning probability is above a certain threshold, they select the outputs corresponding to the input pair $(x,y)=(0,2)$ as the raw keys, followed by the data-reconciliation procedure to obtain the final keys.
One can immediately see that the key rate cannot exceed $\gamma/2$, which is half of the optimal key rate of the MPG-based protocol.

To make a full comparison between the MPG-based protocol and protocols based on the CHSH game, we consider a variety of protocols based on biased CHSH games~\cite{LawsonEtAl:10}.
Suppose that in the CHSH-based protocol, Alice picks $x=0,1$ according to probabilities $1-\varepsilon, \varepsilon$, respectively, while Bob's probabilities of picking $y$ remain the same.
Then, the optimal key rate of the protocol becomes $\gamma(1-\varepsilon)/2$, which is higher than that of the standard CHSH-based protocol and is approaching $\gamma$ when $\varepsilon\rightarrow0$.
We provide the details of the biased CHSH game and its induced DIQKD protocols in \S IIIA of Supplemental Material~\cite{Note1}.

We compare the performance of MPG-based protocol and CHSH-based protocols with different input probabilities.
We suppose that the optimal quantum strategy for the biased CHSH game is used to implement the CHSH-based protocol~\cite{LawsonEtAl:10, AcinEtAl:12}.
It turns out that when introducing the white noise, it is equivalent to treating the distributed state as $\rho_{\nu}$.
When a non-click event occurs, Alice (or Bob) will output $0$ for any input~$x$ (or $y$) such that a deterministic classical strategy is equivalently selected.

The results are shown again in Fig.~\ref{fig:key-rate-visibility-efficiency}, where the key-rate lower bounds of the CHSH-based protocols with different $\varepsilon$s are plotted with blue dashed lines.
These key-rate lower bounds are obtained in a similar fashion to calculating the key-rate lower bound for the MPG-based protocol, as presented in \S IIIB of Supplemental Material~\cite{Note1}.
We observe that all key rates decrease when state visibility or detection efficiency decreases.
As expected, decreasing $\varepsilon$ can increase the optimal key rate of CHSH-based protocols. %, as seen by the values when $\nu=1$ or $\eta=1$.
Nevertheless, these protocols cannot exceed the key rate of the MPG-based protocol in regions of $\nu > 0.978$ and $\eta > 0.982$,
%replace{0.952}{\textbf{0.982}},
showing the advantage of the MPG-based protocol in these regions.

{\em Discussion.}---%%
We have shown that our MPG-based protocol generates more secret keys per instance of protocol than CHSH-based protocols, in regions where the state visibility and detection efficiency are high. % surpass
% This comparison is made on the basis that the costly resource is the number of protocol instances, which in practice relates to the repetition rate of the setup or the time of executing the protocol.
%\add{
This comparison is made on the basis that the costly resource is the number of nonlocal games executed in the experiment~\cite{Gonzales-UretaEtAl:21, PrimaatmajaEtAl:22}.
Indeed, the secrecy of the keys in DIQKD is guaranteed by winning nonlocal games.
Therefore, our result implies that a more complex nonlocal game may lead to a higher amount of secure keys per game round.
%}

%\add{%
Nevertheless, the number of nonlocal games does not necessarily equal to the number of entangled states generated by the source, which is usually the considered resource in practice.
In the case where the source generates a certain number of 2-qubit entangled states, the CHSH-based protocol is preferred.
This is because the CHSH-based protocol can run twice as many rounds as that of a MPG-based protocol, and more keys could be produced.
Also, the CHSH-based protocol is more robust against diminished state visibility and detection efficiency.
%}
%In other words, we have assumed that high-dimensional entangled quantum states can be produced in each protocol instance, as done in recent papers involving multiple input and output Bell inequalities~\cite{Gonzales-UretaEtAl:21}.
%In this sense, our results imply that, when more entanglement is involved in each protocol instance, or when adopting a more sophisticated nonlocal game such as the MPG, the key rate of DIQKD as defined in the above can be improved.
% Nevertheless, the CHSH-based protocols still have advantages in practical situations when the state visibility and the detection efficiency are low.

As for the realization of Protocol~\ref{alg:msg-diqkd-protocol}, the resource state $\Psi_2$ can be produced using two identical preparation of entangled singlets, or using the hyper-entanglement technique to reduce the experimental overheads~\cite{XuEtAl:22a, YangEtAl:05, AolitaEtAl:12}.
An obvious downside of the protocol is its high requirements for state visibility and detection efficiency.
The selection of platforms is important to fulfill the desired requirements.
For instance, the platform with remote matter-qubit entanglement can provide a higher detection efficiency.
Meanwhile, theoretical improvements, including the use of on-maximally entangled states~\cite{WoodheadEtAl:21,SekatskiEtAl:21}, noisy-preprocessing procedures~\cite{HoEtAl:20}, and post-selection techniques~\cite{XuEtAl:22}, can be considered to reduce the requirements of the protocol on the experimental imperfections.

In addition, regarding the higher key rate of the MPG-based protocol over CHSH-based protocols, one may wonder if there are improvements on the CHSH-based protocol such that the key rate $\gamma$ can be achieved.
For instance, in the standard CHSH-based protocol, one can add a key-generation agreement after the state distribution but before the measurements.
Such step allows Alice and Bob to do either key generation or CHSH test [i.e., the original rounds corresponding to $(x,y)=(1,2)$ do not exist], which theoretically enables the key rate to be as high as $\gamma$.
However, the security of the modified protocol requires an additional assumption that no unwanted information is leaked during the key-generation agreement step~\footnote{Otherwise, the untrusted devices may communicate with the adversary in the agreement step of each round}.
We remark that the MPG-based protocol can achieve a key rate as high as $\gamma$ without relying on the above additional assumption.

{\em Conclusions.}---%%
We have proposed the DIQKD protocol based on the MPG and have provided the security analysis of the protocol against the collective attacks in the asymptotic scenario.
We have numerically characterized the regions of two noise parameters, namely, state visibility and detection efficiency, when the MPG-based protocol can produce secure keys.
We have further shown that, in certain regions, the MPG-based protocol has a higher key rate over a variety of protocols based on CHSH games.
Our result shows the advantage of a sophisticated nonlocal game in DIQKD protocols and the potential usage of high-dimensional entanglement in device-independent quantum information tasks.

\begin{acknowledgments}
We gratefully acknowledge valuable discussions with Nai-Le Liu, Kai Chen, Li Li, and Valerio Scarani.
We also thank Enrique Cervero Martín and Marco Tomamichel for pointing out errors in a previous version of this paper and for drawing our attention to related works~\cite{Vidick:2017-Parallel, JainEtAl:2020-Parallel, JainKundu:2023-Direct}, which we had not known.
This work was supported by
National Natural Science Foundation of China (No.~62031024, No.~12005091, No.~12104444),
National Key Research and Development Program of China (No.~2020YFA0309700),
Shanghai Academic/Technology Research Leader (No.~21XD1403800),
and Shanghai Science and Technology Development Funds (No.~22JC1402900).
Y.M.\ acknowledges support from the China Postdoctoral Science Foundation (Grant No. 2021M693093).
F.X.\ acknowledges the support from the Tencent Foundation.
\end{acknowledgments}

\bibliography{Ref_DIQKD}
\end{document}

\title{Supplemental Material:\\Device-Independent Quantum Key Distribution \\
based on the Mermin-Peres Magic Square Game}

\author{Yi-Zheng Zhen}
%\email{yizheng@ustc.edu.cn}
\affiliation{Hefei National Research Center for Physical Sciences at the Microscale and School of Physical Sciences, University of Science and Technology of China, Hefei 230026, China}
\affiliation{Shanghai Research Center for Quantum Science and CAS Center for Excellence in Quantum Information and Quantum Physics, University of Science and Technology of China, Shanghai 201315, China}

\author{Yingqiu Mao}
\affiliation{Hefei National Research Center for Physical Sciences at the Microscale and School of Physical Sciences, University of Science and Technology of China, Hefei 230026, China}
\affiliation{Shanghai Research Center for Quantum Science and CAS Center for Excellence in Quantum Information and Quantum Physics, University of Science and Technology of China, Shanghai 201315, China}

\author{Yu-Zhe Zhang}
\affiliation{Hefei National Research Center for Physical Sciences at the Microscale and School of Physical Sciences, University of Science and Technology of China, Hefei 230026, China}
\affiliation{Shanghai Research Center for Quantum Science and CAS Center for Excellence in Quantum Information and Quantum Physics, University of Science and Technology of China, Shanghai 201315, China}

\author{Feihu Xu}
\email{feihuxu@ustc.edu.cn}
\affiliation{Hefei National Research Center for Physical Sciences at the Microscale and School of Physical Sciences, University of Science and Technology of China, Hefei 230026, China}
\affiliation{Shanghai Research Center for Quantum Science and CAS Center for Excellence in Quantum Information and Quantum Physics, University of Science and Technology of China, Shanghai 201315, China}
\affiliation{Hefei National Laboratory, University of Science and Technology of China, Hefei 230088, China}

\author{Barry C. Sanders}
\email{bsanders@ustc.edu.cn}
\affiliation{Hefei National Research Center for Physical Sciences at the Microscale and School of Physical Sciences, University of Science and Technology of China, Hefei 230026, China}
\affiliation{Shanghai Research Center for Quantum Science and CAS Center for Excellence in Quantum Information and Quantum Physics, University of Science and Technology of China, Shanghai 201315, China}
%\affiliation{Hefei National Laboratory, University of Science and Technology of China, Hefei 230088, China}
\affiliation{Institute for Quantum Science and Technology, University of Calgary, Alberta T2N 1N4, Canada}

\maketitle

\section{Strategies for the MPG}
\label{sec:strategiesMPG}
\subsection{Classical strategies}
\label{subsec:classicalstrategies}
Classical strategies for any nonlocal game can be viewed as a convex combination of deterministic strategies~\cite{BrunnerEtAl:14}, according to which players give a definite answer to any question from the referee.
For the MPG, one deterministic strategy is shown in Table.~\ref{tab:MPGclassical}.
\begin{table}[h]
\centering
(a)
\begin{tabular}{|c|c|c|}
\hline
0 & 0 & 0\\
\hline
0 & 0 & 0\\
\hline
1 & 1 & 0\\
\hline
\end{tabular} 
\qquad
(b)
\begin{tabular}{|c|c|c|}
\hline
0 & 0 & 0\\
\hline
0 & 0 & 0\\
\hline
1 & 1 & 1\\
\hline
\end{tabular}
\caption{\label{tab:MPGclassical}%
One example of the deterministic strategy for the MPG. (a) and (b) are Alice's and Bob's deterministic strategies, respectively.}
\end{table}
It can be seen that Alice and Bob win all grids except the $(3,3)$-grid, which leads to the winning probability of $\nicefrac89$.
By exchanging the rows and/or the columns of the above deterministic strategy simultaneously, or by inverting $0$s and $1$s in the table, one can obtain the other optimal deterministic strategies, each of which can win eight of nine grids.
Therefore, any classical strategy can win the game with the maximal winning probability $\nicefrac89$, which corresponds to the convex combination of optimal deterministic strategies.
Moreover, since Alice's outputs must satisfy $\oplus_{x,i}a^x_{i}=0$ and Bob's outputs must satisfy $\oplus_{y,j}b^y_{j}=1$, any deterministic strategy cannot win the game definitely, and neither does any classical strategy.

\subsection{Quantum strategies}
\label{subsec:quantumstrategies}
Quantum strategies allow players to obtain outputs by measuring quantum states.
When entangled states are shared between players, the measurement outcomes are correlated in a `spooky' way that cannot be simulated by classical strategies.
This feature also allows quantum strategies to win the game over any classical strategy.

For the MPG~\cite{Mermin:90a, Peres:90}, the optimal quantum strategy~\cite{Cabello:01, GisinEtAl:07} has been proven to be a set of measurements on two maximally entangled states $\Psi_{\text{A}_1\text{B}_1}\otimes\Psi_{\text{A}_2\text{B}_2}$.
Here, the measurement set is described by a $3\times3$ table of observables
\begin{table}[!htb]
\centering
\begin{tabular}{|c|c|c|}
\hline & & \\
$I\otimes Z$ & $Z\otimes I$ & $Z\otimes Z$\\
& & \\ \hline & & \\
$X\otimes I$ & $I\otimes X$ & $X\otimes X$\\
& & \\ \hline & & \\
$-X\otimes Z$ & $-Z\otimes X$ & $Y\otimes Y$\\
& & \\ \hline
\end{tabular}
\caption{\label{tab:MPGquantum}%
The measurements of the optimal quantum strategy for the MPG, where $X,Z$ are Pauli-X and Pauli-Z operators, respectively. }
\end{table}
For referee's question~$x$ and $y$, respectively, Alice picks the $x$-row operators and measures $\text{A}_1\text{A}_2$ while Bob picks the $y$-column operators and measures $\text{B}_1\text{B}_2$.
In particular, the eigenvalues $+1$ and $-1$ of each observable in the table corresponds to output $0$ and $1$, respectively.
We emphasize that the observables in each row and each column of the table are jointly measurable, such that three values are obtained in each run of the measurement.
To see this clearly, we express the POVMs corresponding to Alice's and Bob's measurements as $M_{a_0a_1}^x=\ket{m_{a_0a_1}^x}\bra{m_{a_0a_1}^x}$ and $N_{b_0b_1}^y=\ket{n_{b_0b_1}^y}\bra{n_{b_0b_1}^y}$, where
\begin{align}
\Ket{m_{a_0a_1}^{x=0}} &:=X^{a_1}\otimes X^{a_0}\Ket{00},\\
\Ket{m_{a_0a_1}^{x=1}} &:=Z^{a_0}\otimes Z^{a_1}\Ket{++},\\
\Ket{m_{a_0a_1}^{x=2}} &:=Z^{a_0}\otimes Z^{a_1}\frac{\Ket{+1} -\Ket{-0}}{\sqrt{2}},\\
\Ket{n_{b_0b_1}^{y=0}} &:=Z^{b_1}\otimes X^{b_0}\Ket{+0} \\
\Ket{n_{b_0b_1}^{y=1}} &:=X^{b_0}\otimes Z^{b_1}\Ket{0+} \\
\Ket{n_{b_0b_1}^{y=2}} &:=X^{b_0}\otimes Z^{b_1}\frac{\Ket{00} +\Ket{11}}{\sqrt{2}}.
\end{align}
Here, $a_0,a_1,b_0,b_1\in\left\{ 0,1\right\} $, $\ket{0/1}$ are two eigenstates of $Z$ with eigenvalues $+1/-1$, respectively, and $\ket{+/-}$ are two eigenstates of~$x$ with eigenvalues $+1/-1$, respectively.
It can be verified that the above optimal quantum strategy yields the following probability distributions
\begin{align}
P\left(a_{y}^xb_{x}^y=00|xy\right) & =P\left(a_{y}^xb_{x}^y=11|xy\right)=\frac{1}{2},\\
P\left(a_{y}^xb_{x}^y=01|xy\right) & =P\left(a_{y}^xb_{x}^y=10|xy\right)=0,
\end{align}
where $a_{y}^x$ and $b_{x}^y$ denote the overlapped grid of Alice and Bob's outputs when they receive~$x$ and $y$, respectively.

\section{The security of the MPG-protocol beyond the ideal case}
\label{subsec:securityMPG}
\begin{figure}[htbp]
\centering
\includegraphics[width=0.95\columnwidth]{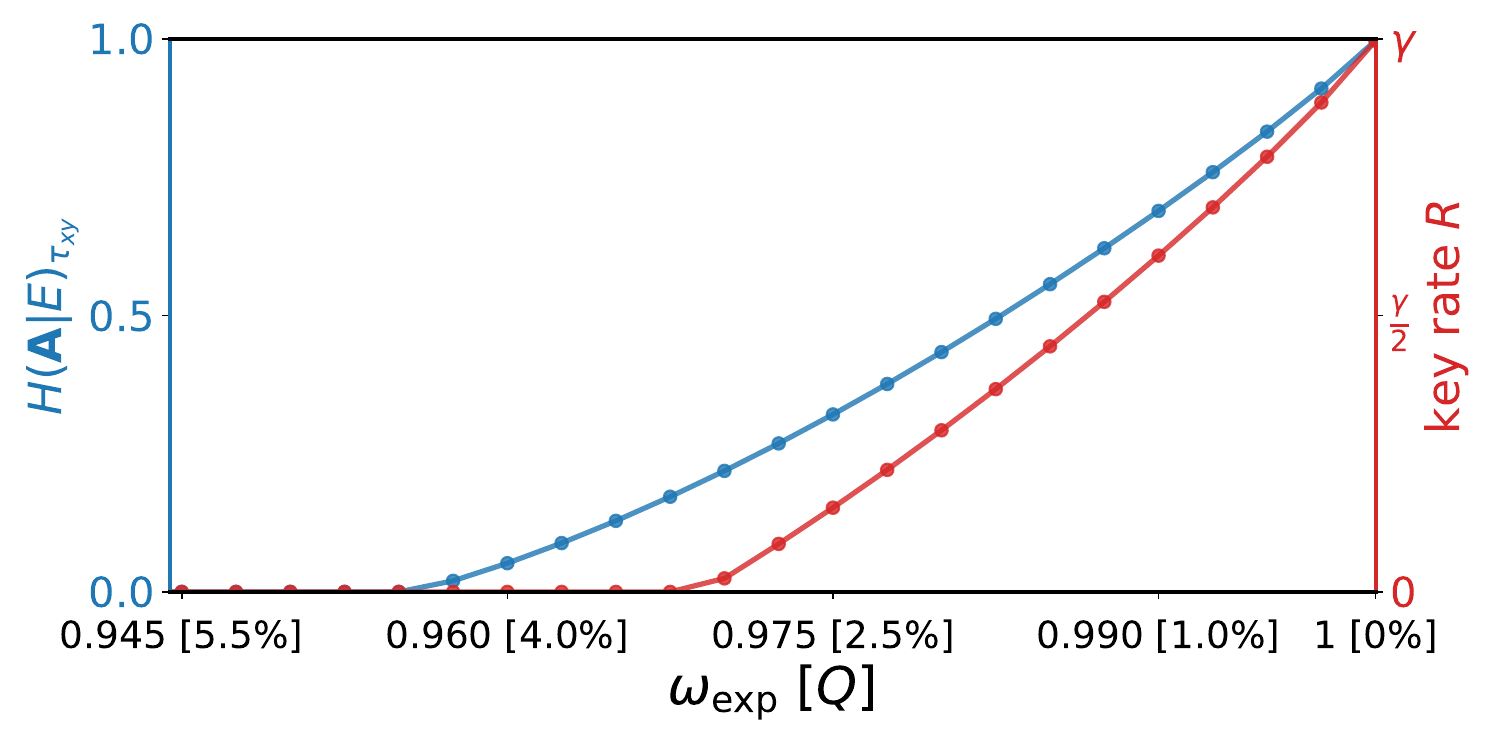}
\caption{\label{fig:hage-key-rate}%
\textbf{The change of the secure randomness produced by the MPG-based protocol, characterized by $H(\boldsymbol{A}|E)_{\tau_{xy}}$, with the expected average winning probability $\omega_\text{exp}$ (blue), and the key rate of an example of the implementation of the protocol changed with the quantum bit error rate $Q$ (red).}}
\end{figure}

Here, we show that the MPG-based protocol can generate secure keys for certain non-ideal cases.
We first calculate the lower bound $H(\boldsymbol{A}|E)_{\tau_{xy}}$ verses the expected winning probability $\omega_\text{exp}$ from Eq.~(11) in the main text.
Because exchanging quantum measurements with different pairs of $(x,y)$ does not change the average winning probability, $H(\boldsymbol{A}|E)_{\tau_{xy}}$ has the same value for all input pairs $(x,y)$ when $\omega_\text{exp}$ is fixed.
The result is shown as the blue line in Fig.~\ref{fig:hage-key-rate}, where we have set $m=16$ and selected NPA hierarchy 2 in the numerical calculation.
We observe that while the lower bound of $H(\boldsymbol{A}|E)_{\tau_{xy}}$ decreases when $\omega_\text{exp}$ becomes smaller, $H(\boldsymbol{A}|E)_{\tau_{xy}}$ has a positive value for all $\omega_\text{exp}>0.9575$. %\replace{0.915}{\textbf{0.9575}}.
This implies that in this region, the MPG-based protocol can produce secure bits for Alice.

To see that the protocol can also produce secure keys between Alice and Bob, we consider a simple implementation of the protocol.
We suppose that the quantum state $\Phi_q = q\Psi_2+(1-q)\mathds{1}_4/4$, where $0\leqslant q\leqslant 1$ and $\mathds{1}_4$ is the identity matrix for the 4-dimensional qudit, is distributed and that the measurements of the optimal quantum strategy for the MPG are used to generate the data.
We set $\omega_\text{exp}=(1+q)/2$ when the distributed state is $\Phi_q$.
Then, we plot the key rate $R$ with the so-called quantum bit error rate, which is defined as $Q=\sum_{xy}\pi(x,y)P(a_{x}\neq b_{y}|xy)$~\cite{PironioEtAl:09} and equals to $(1-q)/2$ in our case, as shown by the red line in Fig.~\ref{fig:hage-key-rate}.
We observe that in this example, $R$ is positive when $Q\apprge3\%$,
%\replace{$6\%$}{\textbf{$3\%$}},
which corresponds to $\omega_\text{exp}\apprge0.97$,
%\replace{0.94}{\textbf{0.97}},
which validates the security of the protocol in the non-ideal case.
In particular, when $q=1$ and $\omega_\text{exp}=1$, we have $R$ approaching $\gamma$, which verifies the optimal case.

\section{The CHSH-based DIQKD protocol}
\label{sec:CHSHprotocol}
\subsection{The biased CHSH game and the corresponding DIQKD protocol}
\label{subsec:CHSHgame}

The CHSH game is performed as follows~\cite{BrunnerEtAl:14}.
In each round, the referee randomly picks $x,y\in\{0,1\}$ to Alice and Bob according to the probability distribution $\pi(x,y)$, and Alice and Bob reply with $a,b\in\{0,1\}$, respectively.
During the game, Alice and Bob are forbidden to communicate with each other.
The winning condition is $a\oplus b=xy$.
After many rounds, the referee decides whether Alice and Bob win the game according to the average winning probability
\begin{equation}
I=\sum_{x,y}\pi\left(x,y\right)P\left(a\oplus b=xy|xy\right),\label{eq:biased-chsh}
\end{equation}
where $P(a\oplus b=xy|xy)$ is the probability that Alice and Bob win for each input pair $(x,y)$.

The standard CHSH game has the unbiased input; i.e., $\pi(x,y)=1/4$ for all $(x,y)$.
It is well-known that the optimal classical strategy leads to the maximal average winning probability $3/4$, while the optimal quantum strategy can achieve $(2+\sqrt{2})/4$~\cite{BrunnerEtAl:14}.
However, when $\pi(x,y)$ is not uniformly distributed, the optimal average winning probabilities from classical and quantum strategies will change~\cite{LawsonEtAl:10}.
In the main text, we are interested in the biased CHSH game where $\pi(0,0)=\pi(0,1)=(1-\varepsilon)/2$ and $\pi(1,0)=\pi(1,1)=\varepsilon/2$.
The average winning probability, in the form of Eq.~\eqref{eq:biased-chsh}, can be equivalently written as
\begin{equation}
I_{\varepsilon}=\frac{1-\varepsilon}{2}\left(E_{00}+E_{01} \right)+\frac{\varepsilon}{2}\left(E_{10}-E_{11}\right),
\end{equation}
where
\begin{equation}
E_{xy}:=\sum_{a_{x}b_{y}}(-1)^{a_{x}+b_{y}}P(a_{x}b_{y}|xy).
\end{equation}
In fact, $I_{\varepsilon}$ is equivalent to the asymmetric CHSH inequality introduced in~\cite{AcinEtAl:12}, which has a classical upper bound $1-\varepsilon$.
Thus, any classical strategy can win the biased CHSH game with a winning probability as large as $1-\varepsilon$. 
The optimal quantum strategy can exceed this threshold with quantum measurements
\begin{subequations}\label{eq:biased-chsh-meas}
\begin{align}
 A_0 &= Z, & B_0 &= \cos\mu Z + \sin\mu X, \\
 A_1 &= X, & B_1 &= \cos\mu Z - \sin\mu X,
\end{align}
\end{subequations}
on the maximally entangled two-qubit state $\Psi^+$, where $\mu$ satisfies $\tan\mu=\varepsilon/(1-\varepsilon)$~\cite{AcinEtAl:12}.

We provide the DIQKD protocol based on the biased CHSH game in Protocol~\ref{alg:chsh-diqkd-protocol}, which is based on protocols investigated in~\cite{AcinEtAl:07, PironioEtAl:09}.

\begin{algorithm}[htbp]
\centering
\caption{\label{alg:chsh-diqkd-protocol}%
The CHSH-based DIQKD protocol}
\begin{description}
 \item [Input] $N$ -- number of rounds,
 $\varepsilon$ -- probability that Alice picks key generation basis,
 $I_\text{exp}$ -- expected winning probability of the CHSH game.
 \item [Output] $\boldsymbol{K}_{\text{A}}$ -- Alice's final key, $\boldsymbol{K}_\text{B}$ -- Bob's final key.
\end{description}\vspace{1mm}
\begin{description}
\item [Data generation]  In each round, Alice randomly picks $x\in\{0,1\}$ with probabilities $1-\varepsilon, \varepsilon$, and Bob randomly picks $y\in\{0,1,2\}$ with probabilities $(1-\gamma)/2, (1-\gamma)/2, \gamma$.
They inject $x$ and $y$ to their devices and record the outputs $a_x$ for Alice and $b_y$ for Bob, respectively.

\item [Announcement and sifting] Alice and Bob announce their inputs of all the rounds.
They select the rounds corresponding to $(x,y)\in\{0,1\}^2$ as test rounds, and the rounds corresponding to $(x,y)=(0,2)$ as key rounds.

\item [Parameter estimation] Alice and Bob announce the outputs of test rounds, from which they estimate the average winning probability $I$ of the CHSH game.
If $I<I_\text{exp}$, they abort the protocol;
otherwise, they record the output of key rounds as raw keys $\boldsymbol{A}$ and $\boldsymbol{B}$, respectively.

\item [Data reconciliation] Alice and Bob apply error correction and privacy amplification on the raw keys $\boldsymbol{A}$ and $\boldsymbol{B}$ to obtain final secure keys $\boldsymbol{K}_\text{A}$ and $\boldsymbol{K}_\text{B}$, respectively.
\end{description}
\end{algorithm}

\subsection{The protocol and the key rate}
\label{subsec:CHSHprotocol-KeyRate}

Following the same definition of key rate in the main text, we express the key rate of Protocol~\ref{alg:chsh-diqkd-protocol} as
\begin{equation}\label{eq:key-rate-chsh}
R=  \min\left\{ 0, \frac{\gamma(1-\varepsilon)}{2}\inf_{\left(\rho,{\cal M}\right)}
r\left(\tau^{\prime}\right)\right\},
\end{equation}
subject to
\begin{equation}\label{eq:constr-chsh}
I\left(\rho,{\cal M}\right)\geqslant I_\text{exp}.
\end{equation}
Here, $\gamma(1-\varepsilon)/2$ is the ratio between the number of raw keys and the number of all rounds, $(\rho,{\cal M})$ stands for quantum strategies for the biased CHSH game, $I(\rho,{\cal M})$ is the average winning probability of that strategy, and $r(\tau^{\prime})$ is defined in Eq.~(6) of the main text with
\begin{equation}\label{eq:ccq-chsh}
    \tau^{\prime} := \sum_{a,b} \Ket{a}\Bra{a}_\text{A} \otimes \Ket{b}\Bra{b}_\text{B} \otimes \hat{\phi}_\text{E}(ab|x=0,y=2),
\end{equation}

Similar to the calculation of the key-rate lower bound of the MPG-based protocol in the main text, the key rate of the CHSH-based protocol [Eq.~\eqref{eq:key-rate-chsh}] can be bounded as follows.
The $H(\boldsymbol{A}|E)_{\tau^{\prime}}$ term in the expression of $r(\tau^{\prime})$ has a lower bound
\begin{equation}\label{eq:HAgElowerbound-chsh}
H\left(\boldsymbol{A}|E\right)_{\tau^{\prime}}\geqslant c_{m}+\sum_{k=1}^{m-1}c_{k}\min\braket{\psi|G^{\prime}|\psi},
\end{equation}
where the minimization is over all possible quantum state $\ket{\psi}=\ket{\psi}_\text{ABE}$ and measurements ${\cal M}=\{ \{M^{x}_a\}_{x}, \{N^{y}_b\}_{y} \}$, subject to
\begin{subequations}
\begin{align}\label{eq:SDP-CHSH}
    G^{\prime} &:= 
    \sum_{a\in\left\{ 0,1\right\} }\left\{ M^{x=0}_a\left[Z_a+Z_a^{\dagger} + \left(1-t_{k}\right)Z_a^{\dagger}Z_a\right]\right.\nonumber\\ 
    & \hspace{2cm} \left.+t_{k}Z_aZ_a^{\dagger}\right\},\\
    I_{\text{exp}} &\leqslant I\left(\rho,{\cal M}\right),
    \quad \rho = \text{tr}_{\text{E}}\left[\ket{\psi}\bra{\psi}_{\text{ABE}}\right],\\ 
    0 &= \left[M_a^{x},N_b^{y}\right]= \left[M_a^{x},Z\right] = \left[Z_a,N_b^{y}\right],\\
    & \forall a,b,x,y\in\{0,1\}.
\end{align}
\end{subequations}
By resolving the above minimization with the NPA hierarchy~\cite{NavascuesEtAl:07} and its corresponding semi-definite program~\cite{BrownEtAl:21a}, the lower bound of $H(\boldsymbol{A}|E)_{\tau^{\prime}}$ is obtained.
The other term in $r(\tau^{\prime})$, namely, $H(\boldsymbol{A}_0|\boldsymbol{B}_2)$
can be estimated from the outputs of key rounds.
The difference between the above two values is indeed a lower bound for the key rate.

To calculate the $H(\boldsymbol{A}|\boldsymbol{B})$ term in $r(\tau^{\prime})$, we suppose that the protocol is implemented by quantum measurements used are $A_{0,1}, B_{0,1}$ described by Eq.~\eqref{eq:biased-chsh-meas} and $\text{B}_2=Z$ on quantum state $\Psi^+$ (before entering noisy channels).

\section{The full-statistic variation of the MPG-based protocol}

As an additional supplementary result, we ask if the full-statistic variation of the protocol can be better than the game-based protocol as we introduced here.
We found that the full-statistic variation is indeed better than the game-based one, although the advantage is slight (see the figure below).

\begin{figure}[h]
    \centering
    \includegraphics[width=0.9\columnwidth]{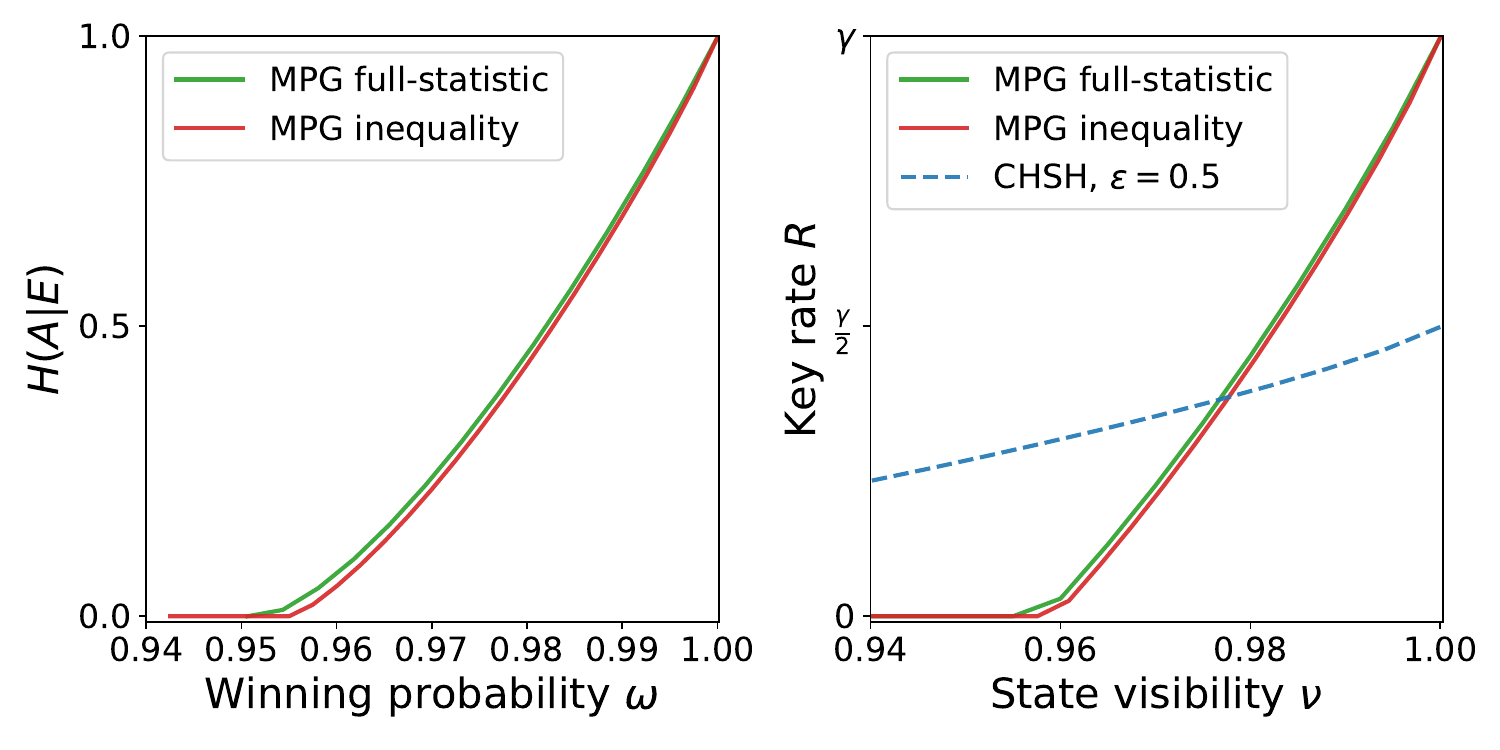}
    \caption{The comparison between our protocol (red) and the full-statistic variation of the protocol (green). (Left) Eve's information on Alice's raw keys with respective to different winning probabilities. (Right) The key rate for different state visibility.}
\end{figure}

To obtain the plots of the full-statistic variation in the above figure, we have replaced the test in the parameter estimation of our protocol with a test that the observed statistics is in agreement with the probabilities obtained by quantum states $\rho_{\nu}\otimes\rho_{\nu}$ where $\rho_{\nu}=\nu\Psi^++(1-\nu)\mathds1/2$.

\bibliography{Ref_DIQKD}